\documentstyle[epsf,fleqn,12pt]{article}

\advance\textheight by \topskip
\newcommand{\sa}{self-adjoint}
\newcommand{\se}{Schr\"odinger equation}
\newcommand{\bm}{ Bohmian mechanics}

\newcommand{\wf}{wave function}

\newcommand{\bd}{\begin{displaymath}}
\newcommand{\ed}{\end{displaymath}}
\newcommand{\be}{\begin{equation}}
\newcommand{\ee}{\end{equation}}
\newcommand{\bi}{\begin{itemize}}
\newcommand{\ei}{\end{itemize}}
\newcommand{\ba}{\begin{eqnarray}}
\newcommand{\ea}{\end{eqnarray}}

\def\biggi{\bigskip\noindent}

\newcommand{\lit}{ \lim_{t \to \infty} }
\newcommand{\slim}{ {\mbox{\rm s-}} \lim}
\newcommand{\lir}{ \lim_{R \to \infty} }

\newcommand{\psihat}{\hat{\psi}}

\newcommand{\xt}{\frac{\bx}{t}}

\newcommand{\xyt}{\frac{\bx\cdot\by}{t}}
\newcommand{\xxt}{\frac{x^2}{2t}}
\newcommand{\yyt}{\frac{y^2}{2t}}
\newcommand{\kkt}{\frac{k^2 t}{2}}

\newcommand{\bj}{{\bf j}}

\newcommand{\bx}{{\bf x}}
\newcommand{\by}{{\bf y}}

\newcommand{\bv}{{\bf v}}
\newcommand{\bn}{{\bf n}}
\newcommand{\bk}{{\bf k}}

\newcommand{\PKR}{{\partial B_R}}
\newcommand{\jpsi}{\bj^{\psi_t}}

\newcommand{\SIC}{scattering-into-cones-theorem}
\newcommand{\FAS}{flux-across-surfaces-theorem}

\newtheorem{satz}{Satz}[section]

\newtheorem{definition}[satz]{Definition}
\newtheorem{remark}[satz]{Remark}
\newtheorem{example}[satz]{Example}

\def\IR{\mathchoice{ \hbox{${\rm I}\!{\rm R}$} }
                   { \hbox{${\rm I}\!{\rm R}$} }
                   { \hbox{$ \scriptstyle  {\rm I}\!{\rm R}$} }
                   { \hbox{$ \scriptscriptstyle  {\rm I}\!{\rm R}$} }   }
\def\IP{\mathchoice{ \hbox{${\rm I}\!{\rm P}$} }
                   { \hbox{${\rm I}\!{\rm P}$} }
                   { \hbox{$ \scriptstyle  {\rm I}\!{\rm P}$} }
                   { \hbox{$ \scriptscriptstyle  {\rm I}\!{\rm P}$} }   }
\def\IE{\mathchoice{ \hbox{${\rm I}\!{\rm E}$} }
                   { \hbox{${\rm I}\!{\rm E}$} }
                   { \hbox{$ \scriptstyle  {\rm I}\!{\rm E}$} }
                   { \hbox{$ \scriptscriptstyle  {\rm I}\!{\rm E}$} }   }

\font\sqi=cmssqi8

\def\ICi{\hbox{$C\hskip-.7em\raise.75pt\hbox{\sqi I}\;\;$}}

\begin{document}

\noindent
Contribution to ``Bohmian Mechanics and Quantum Theory:
An Appraisal,''\\
edited by J.T.\ Cushing, A.\ Fine, and S.\ Goldstein

\bigskip

\biggi
{\huge \bf Scattering Theory}

\medskip\noindent
{\huge \bf  from a Bohmian Perspective}

\bigskip

\biggi
Martin Daumer\\
Mathematisches Institut der Universit\"{a}t M\"{u}nchen,\\
Theresienstra{\ss}e 39,
80333 M\"{u}nchen, Germany

\biggi
\today

 Quantum mechanical scattering theory is a subject with a long and winding
history. We shall pick out some of the most important concepts and ideas
of scattering theory and look at them
{}from the perspective of Bohmian mechanics: \bm , having real particle
trajectories,
provides an excellent basis for analyzing scattering phenomena.

\section{A very brief historical sketch}

We begin with a quote taken from Born 1926, shortly after Heisenberg 1925
had invented matrix mechanics and Schr\"odinger  1926 his wave mechanics:

\bigskip

{\footnotesize
Neither of these two conceptions appear satisfactory to me. I should like
to
attempt here to give a third interpretation and to test its utility on
collision processes.
In this attempt, I adhere to an observation of Einstein on the
relationship of wave field and light quanta; he said, for example, that
the waves are
present
only to show the corpuscular light quanta the way, and he spoke in the
sense of
a ``ghost field''. This determines the probability that
a light quantum \dots takes a certain path; \dots
And here it is obvious to regard the de Broglie-Schr\"odinger waves as
the ghost
field or, better, ``guiding field''.

I should therefore like to investigate experimentally the following
idea: the guiding
field, represented by a scalar function $\psi$ of the coordinates of
all
the particles
involved and the time, propagates in accordance with Schr\"odinger's
differential
equation. \dots
The paths of these corpuscules are determined only to the extent that
the laws
of energy and momentum restrict them; otherwise, only a probability
for a certain path is found.}

\bigskip

and the

\bigskip

{\footnotesize
Closing remarks:
On the basis of the above discussion, I should like to put forward
the opinion
that quantum mechanics permits not only the formulation and solution
of the
problem of stationary states, but also that of transition processes.
In these
circumstances Schr\"odinger's version appears to do justice to the
facts in
by far the easiest manner; moreover, it permits the retention of the
conventional ideas of space and time in which events take place in a
completely
normal manner. On the other hand, the proposed theory does not
correspond to the
requirement of the causal determinacy of the individual event. In
my preliminary
communication I stressed this indeterminacy quite particularly,
since it appears
to me in best agreement with the practice of the experimenter.
But it is natural for him
who will not be satisfied with this to remain unconverted and to
assume that
there are other parameters, not given in the theory, that
determine the individual
event. In classical mechanics these are the ``phases'' of the
motion, i.e.\
the coordinates of the particles at a given instant. It appears
to me {\it a priori}\/
improbable that quantities corresponding to these phases can
easily be introduced into
the new theory, but Mr.~Frenkel has told me that this may
perhaps be the case.
However this may be, this possibility would not alter anything
relating the
practical indeterminacy
of collision processes, since it is in fact impossible to
give the values of the
phases; it must in fact lead to the same formulae as
the ``phaseless'' theory
proposed here.

{\hfill Born 1926, translated in Ludwig 1968, p.224}
}

\bigskip

Schr\"odinger 1926 had shown with his wave mechanics
 how to obtain the discrete
energy levels of the hydrogen atom, by seeking stationary
square integrable solutions
of the \se\ (in units $\hbar=m=1$)
\be
\label{SCHREQ}
i \frac{\partial \psi_t(\bx)}{\partial t}  =
-\frac{1}{2}\Delta \psi_t(\bx) +
V(\bx) \psi_t(\bx).
\ee
What are called stationary states are  solutions
of the form  $\psi_t(\bx) = e^{-iEt}
\psi(\bx)$, where
 $\psi(\bx)$ obeys the stationary \se
\be
\label{STATSCH}
-\frac{1}{2} \Delta \psi(\bx) + V(\bx)\psi(\bx)  =
E\psi(\bx),
\ee
which has square integrable solutions only for certain discrete energies
$E_n < 0$, in agreement with the experimental results.
The physical meaning of the wave function was however unclear.

A description of a time-dependent scattering process
was  soon  given by
 Max Born 1926, who  explored the hypothesis that the wave function
might be a ``guiding field'' for the motion of the electron. As a
consequence
of this hypothesis, Born is led in his paper to the ``statistical
interpretation'' of the wave function: $\rho_t(\bx) = |\psi_t(\bx)|^2$
is the
probability density for a particle to be at point $\bx$ at time $t$.
It follows from (\ref{SCHREQ}) that there is a conserved flux
corresponding
to this density, the quantum flux $\jpsi(\bx) :=
  {\rm Im} \bigl(\psi_t^*(\bx) \nabla \psi_t(\bx)\bigr)$, which
obeys the continuity
equation
\be
\label{CONTEQ}
\frac{\partial |\psi_t(\bx)|^2}{\partial t}  + \nabla \cdot \jpsi(\bx)
= 0.
\ee
Born interpreted the quantum flux  as a probability current of particles.
His basic ansatz,  sometimes called ``naive'' scattering theory
(Reed, Simon 1979, p.\ 355),
is to seek non-normalized solutions of the stationary \se\ (\ref{STATSCH})
for {\it positive} energies $E=k^2/2$, which have  the long-distance
behavior
\be
\label{LONGDIS}
\psi(\bx) \stackrel{x\to\infty}{\sim} e^{i\bk\cdot\bx} + f(\theta,\phi)
\frac{e^{ikx}}{x}.
\ee
$e^{i\bk\cdot\bx}$ is interpreted as the  incoming plane wave and
$f(\theta,\phi) \frac{e^{ikx}}{x}$ as the outgoing spherical wave with
angular dependent density.
The flux corresponding to the incoming wave is $\bk$, such that the
number
of crossings per unit time and a unit surface orthogonal to $\bk$ is $k :=
|\bk|$. The flux
corresponding to the spherical wave is $\frac{\bx}{x^3} k|f(\theta,\phi)|^2$
and is obviously
purely radial. The number
of crossings of the surface element $x^2d\Omega$ of a distant sphere
in a direction
specified by the angles $\theta,\phi$ per unit time divided by the number
of incoming particles  per unit time and unit surface
 is called ``differential cross section,''  and one finds
\be
\label{BORNFORM}
\frac{d\sigma}{d\Omega} = |f(\theta,\phi)|^2.
\ee

This description of a scattering process is however not  convincing for
simple reasons: there is no hint in the equations what
the  particles do which are responsible for
the flux and how their motion is related
to the wave function such that $\rho_t(\bx) = |\psi_t(\bx)|^2$
holds for all times.
Moreover, the picture is entirely time-independent although a scattering
process is certainly a process in space and time; stuff moves.
The arguments leading to
the formula (\ref{BORNFORM}) for the cross section
``wouldn't convince an educated first grader'' (Goldberger, cited in
Simon 1971, p 97).

{}From a physical point of view it might have seemed natural
that a time-dependent justification of Born's time-independent method
was developed, involving a detailed analysis of the behavior
of the \wf\ and its corresponding flux, but scattering theory proceeded
along a
 different direction. (For the reasons why Born later abandoned the idea
of a guiding field  see e.g.\ Beller 1990.)

Heisenberg  aimed at casting all in-principle-measurable quantities,
such as
energy levels and the cross section, into a single abstract object: the
unitary $S$-matrix, or better, writing $S=e^{i\eta}$, into the \sa\
 ``phase matrix''
$\eta$ (Heisenberg 1946).
$S$ should be the map from freely evolving
``in''-states, which are controllable
by an experimenter, to the freely evolving ``out''-states, whose
properties can
be measured.
Motivated by this idea M{\o}ller
introduced the concept of ``wave operators'' (M{\o}ller 1946),
the building blocks of the $S$-matrix: one expects that
in a scattering process any wave function has a simple asymptotic
behavior for large negative and positive times, where it should evolve
almost freely with
$e^{-iH_0t}$ ($H_0 := -\frac{1}{2}\Delta$).
Hence there should be states $\phi_+$ and $\phi_-$ such that
\be
\label{WAVEOPDER}
\lim_{t\to\pm\infty} \|e^{-iHt}\psi - e^{-iH_0t} \phi_{\pm}\|_2 = 0,
\ee
where $\|\cdot\|_2$ is the norm in $L^2$.
If the wave operators
\be
\label{WAVEOPDEF}
\Omega_{\mp} = \slim_{t\to\pm\infty} e^{iHt}e^{-iH_0t}
\ee
(``$\slim$'' denotes the strong limit) exist, then,
if we now imagine that $\phi_\pm$ are given,
obviously $\psi:=\Omega_\mp \phi_\pm$ obeys (\ref{WAVEOPDER}).
(The strange sign convention is a tradition, see Reed, Simon 1979, p.17.)

The operators $\Omega_\pm$ should be unitary (we assume here that there
are no bound states) such that $\Omega_\pm^\dagger = \Omega_\pm^{-1}$.
The $S$-matrix can now be  defined as $S= \Omega_-^\dagger\Omega_+$,
because it  maps freely evolving ``in''-states  onto ``out''-states.

There has been a lot of work on the precise definition and
properties of wave operators.
We
should mention here   the
concept of modified wave operators which was introduced (Dollard 1964)
in order to bypass the problem, that the usual wave operators don't
exist for long range potentials such as the Coulomb potential.
The program of ``asymptotic completeness,'' which aims at proving
certain additional properties of the wave operators and the spectrum
of the Hamiltonian, kept mathematicians and physicists busy for
a long time, until recently this problem could be solved in
great generality (Derezi\'{n}ski, Gerard 1993).

Interestingly enough, these achievements
would not have been possible without
the ``renaissance'' of geometrical ideas (Ruelle 1969, Enss 1978), namely
by realizing that the wave packets evolve in space and time and are
far away from the scattering center most of the time,
 instead of working with very abstract
and complicated methods in momentum space (Faddeev 1965), where this
simple
geometrical picture easily gets lost.

Most of this work on the $S$-matrix, wave operators and asymptotic
completeness
is however not much concerned with the original question of scattering
theory,
the justification of the formula for the cross section (\ref{BORNFORM})
of the ``naive''
scattering theory, which was
commonly used to do the actual numerical calculations.
It is true that formulas for the differential
cross section have been suggested from an analysis of
$\langle \bk'|S|\bk\rangle$,
which was interpreted as the ``probability density''
to find the momentum eigenstate
$|\bk'\rangle$ in the final state $S|\bk\rangle$
(Lippmann and Schwinger 1950
used
 ``adiabatic switching,''  Gell-Mann and Goldberger 1953
Abelian limits), but these arguments
were not much more convincing than the original argument to arrive at
 (\ref{BORNFORM}) (see also Reed, Simon 1979, p.356).

An exception is the work of Ikebe 1960, who
rigorously established the
physicist's notation of expansions in
continuum eigenfunctions (for ``Ikebe''-potentials)
and linked wave operators with solutions of the
``Lippmann-Schwinger equation.''
The time-dependent \wf\ (in physicists notation formally
$\langle\bx|e^{-iHt}|\psi\rangle =
\int d^3k e^{-i\kkt} \langle\bx\widetilde{|\bk\rangle}
\widetilde{\langle \bk|}\psi\rangle)$
may be written as
\be
\label{PSIT}
\psi_t(\bx) = (e^{-iHt}\psi)(\bx) =
(2\pi)^{-3/2}\int d^3k e^{-i\kkt}\phi(\bx,\bk) \psi^{\#}(\bk),
\ee
where the ``generalized eigenfunctions'' $\phi(\bx,\bk)$
are solutions of the
Lippmann-Schwinger equation
\be
\label{LS1}
\phi(\bx,\bk) = e^{i\bk\cdot\bx} - \frac{1}{2\pi} \int d^3y
\frac{e^{-i{k|\bx-\by|}}}{|\bx-\by|}
V(\by) \phi(\by,\bk).
\ee
Here $\psi^{\#}(\bk) :=  (2\pi)^{-3/2} \int d^3x \phi^*(\bx,\bk) \psi(\bx)$
 is the ``generalized Fourier transform'' of $\psi$
and is connected with the wave operators by
$(\widehat{\Omega_-^\dagger \psi})(\bk) = \psi^{\#}(\bk)$,
where $\widehat{\ }$ denotes the usual Fourier transform.
(There is another set of eigenfunctions corresponding to $\Omega_+$
which are in fact the ones most frequently  used.)
The solutions of the Lippmann-Schwinger equation
are also solutions of the stationary \se\
with the asymptotic behavior as in (\ref{LONGDIS})
(the ones for $\Omega_+$) such that
the differential cross section
may be read off their long distance asymptotics as in (\ref{BORNFORM}).
At least one could now find the formula (\ref{BORNFORM}) of the ``naive''
scattering theory
somewhere in an appropriate expansion of the
time-dependent wave function.

Another exception is the work of Dollard 1969.
Dollard suggested to use the probability to find a particle in the far
future
in a given cone $C\subset\IR^3$ as
a natural time-dependent {\it definition} of the
cross section.
 Dollard's ``\SIC ''  relates this probability to the wave operators:
\be
\label{SICT}
\lit \int_C d^3x |\psi_t(\bx)|^2 = \int_C d^3v
|\widehat{\Omega_-^\dagger\psi}(\bv)|^2.
\ee
The \SIC\ has come to be regarded as the fundamental result from
which the differential
cross section ought to be derived (e.g.\ Reed, Simon 1979, p.356, and
Enss, Simon  1980).

Dollard's approach was however criticized
by Combes, Newton and Shtokhamer 1975. They observe that
the experimental relevance of the \SIC\ rests on the connection of
the probability of
finding the particle in the far future in a cone with the probability
that the particle
has, at some time, crossed a given distant surface subtended by the cone.
Heuristically,
the last
 probability should be given by integrating the
quantum mechanical flux over the total time interval and this surface.
(The flux is often used that way in  textbooks.) Combes, Newton and
Shtokhamer hence
conjecture the ``\FAS''
\be
\label{FAST}
\lir \int_0^\infty dt \int_{C\cap\PKR} \bj^{\psi_t} \cdot \bn d\sigma=
\int_C d^3v |\widehat{\Omega_-^\dagger\psi}(\bv)|^2,
\ee
where $B_R$ is the ball with radius $R$ and outward normal $\bn$.
There exists no proof of this theorem. Even the
 ``free \FAS\,'' for freely evolving $\psi_t$,
\be
\label{FASFREET}
\lir \int_0^\infty dt \int_{C\cap\PKR} \bj^{\psi_t} \cdot \bn d\sigma=
\int_C d^3v |\psihat(\bv)|^2
\ee
which should be physically good enough, because the scattered wave packet
is expected to move almost freely
after the scattering is essentially completed,
has not been proven.

This is certainly strange because the physical
importance of
(\ref{FAST}) and (\ref{FASFREET}) for scattering theory
is obvious and the mathematical
problem does not
seem to be too hard.
Perhaps it was  the vagueness in the meaning attached to  the flux which
is responsible for this matter of fact. For example,
the authors try to
reformulate the problem in operator language and are faced with the
problem
that for general $L^2$-functions the current across a given surface
may well be
infinite. Instead of using smooth functions they use ``smeared-out''
surfaces and
therefore fail to find a proof of the original theorem.
Or, for example, they argue that

{\footnotesize
At large distances the
scattering part of the wave function contains outgoing particles only.
Therefore the particles cannot describe loops there and the flux can
be measured by the
interposition of counters on $\PKR$.}\\

and seem to have in mind a picture very
similar to Born's original proposal (cf.\ the Born quote), also without
giving a precise guiding law  for the trajectories which would allow,
for example,
to check the
``no-loop-conjecture.''

Next we want to show how the ideas of Combes, Newton and Shtokhamer
arise naturally
by  analyzing a scattering process in the framework of Bohmian mechanics
(Bell 1987,
Bohm 1952, Bohm and Hiley 1993, D\"urr, Goldstein and Zangh\a'{\i} 1992
and their
essay in this volume, Holland 1993):
in this theory
{\it particles  move} along trajectories determined by the quantum flux,
controlling the expected number of particles crossings of surfaces.
We will sketch the main ideas of the proof of (\ref{FASFREET})
(for the complete proof see  Daumer 1995)
and indicate the extension to the interacting case.
We shall see
that the  \FAS\
in \bm\ is a relation between the flux across a distant surface and the
asymptotic probability of outward crossings of the trajectories
of this surface---obviously the quantity of interest for the scattering
analysis
of {\it any} mechanical theory of point particles.

\section{Bohmian Mechanics}

\bm\ does what not only Born found ``a priori improbable,'' namely it
shows that the
introduction of additional parameters into the theory, represented by
the \se\
(\ref{SCHREQ}),
 which ``determine the
individual event'' is easily possible (see the
essay of D\"urr, Goldstein and Zangh\a'{\i}):
the integral curves of the velocity field
\be
\bv^{\psi_t}(\bx) = \frac{\jpsi}{\rho_t}(\bx) =
{\rm Im}\frac{\nabla \psi_t}{\psi_t}(\bx)
\ee
 which are solutions of
\begin{equation}
\label{BM1}
\frac{d}{dt} {\bx}(t) = {\bf v}^{\psi_t}({\bx}(t)),
\end{equation}
together with an initial position $\bx_0$ determine
the trajectory of the particle.
(For a proof of the global existence of the solutions
for general $N$-particle systems
and a large class of potentials see the essay of Berndl.)
The initial position is distributed according to the quantum equilibrium
probability
$\IP^\psi$  ($\psi$ is normalized) with density $\rho=|\psi|^2$ (for
a justification
of ``quantum equilibrium'' see
 D\"urr, Goldstein and Zangh\a'{\i} 1992).

Thus, in \bm\  a particle moves along a trajectory
guided by the particle's \wf .
Hence, given $\psi_t$, the solutions ${\bx}(t,{\bx}_0)$ of equation
(\ref{BM1})
are random trajectories,
where the randomness comes from the $\IP^\psi$-distributed random initial
position ${\bx}_0$,
$\psi$ being the initial \wf .

Consider  now a region $G\subset \IR^3$ and  let $N^{\Sigma,\Delta}$ be
the  number of crossings of $\bx(t)$ of
subsets  $\Sigma \subset \partial G$ in
time intervals $\Delta \subset [0,\infty)$.
Splitting $N^{\Sigma,\Delta} =: N_+^{\Sigma,\Delta} + N_-^{\Sigma,\Delta}$,
where $N_+^{\Sigma,\Delta}$
denotes the number of outward crossings
and $N_-^{\Sigma,\Delta}$ the number of backward crossings of $\Sigma$ in
$\Delta$,
we define for the number of ``signed crossings''
$N_s^{\Sigma,\Delta} := N_+^{\Sigma,\Delta} - N_-^{\Sigma,\Delta}.$
By the very meaning of the probability flux it is rather
clear (and it can easily be computed, Berndl 1995),
that the expectated value  of these numbers of
crossings in quantum equilibrium is given by integrals of the current,
namely
\ba
\label{EXPECT}
\IE^\psi(N^{\Sigma,\Delta }) &=&
\int_\Delta dt\int_\Sigma |\jpsi \cdot \bn| d\sigma \/ \Bigl( =
\int d^3x |\psi(\bx)|^2
N^{\Sigma,\Delta}(\bx)\Bigr)\\
\label{EXPECT1}
\IE^\psi(N_s^{\Sigma,\Delta}) &=& \int_\Delta dt\int_\Sigma \jpsi
\cdot \bn d\sigma.
\ea
(This relation between the current and the expected number of crossings is
also one of the fundamental insights used in the proof of global existence
of solutions.)

\section{Scattering analysis of Bohmian mechanics}

We want to analyze the {\it scattering regime} of Bohmian mechanics,
i.e.\ the
asymptotic behavior of the
distribution of crossings of the trajectories traversing some distant
surface surrounding the
scattering center (see also Daumer 1995).

As surfaces we chose, for the sake of simplicity, spheres and we fix
the  notation
illustrated in  figure 1.
(We may imagine the surface surrounding the scattering center or,
more generally,
simply an area in which the particle happens to be.)

We consider the random variables
(functions of the paths) {\it first exit time} from $B_R$
\be
\label{ESCAPETIME}
t_e := \inf \{t\ge 0| \bx(t) \notin B_R\}
\ee
and the corresponding {\it exit position}
\be
\label{ESCAPEPOSITION}
\bx_e = \bx(t_e).
\ee
Upon solving (\ref{BM1}) and (\ref{SCHREQ}) the statistical
distributions for $t_e$ and $\bx_e$
can of course be calculated (see e.g.\ Leavens 1990 and his
essay on the
related problem of tunneling times).
In general we should expect this to be a very hard task
but it turns out that if $\PKR$ is at most crossed once by every
trajectory---this is
what we expect to happen asymptotically in the scattering
regime---a very simple formula
involving the current obtains.

The probability of the exit positions $\IP^\psi(\bx_e \in R\Sigma)$
should become ``independent''
of $R$ for large $R$ such that we may focus on the map
$\sigma^\psi: {\cal B}(S^2) \to \IR^+$ defined by
\be
\label{SIGMA0}
\sigma^\psi(\Sigma) := \lir \IP^\psi(\bx_e\in R\Sigma) =
\lir \IP^\psi(\frac{\bx_e}{x_e} \in \Sigma),
\ee
where  $\frac{\bx_e}{x_e}$   is the exit direction, which we expect
to be a probability
measure on
the unit sphere (if eventually all trajectories go off to infinity,
as they should.)
This measure
gives us the asymptotic probability of outward crossings of a distant
surface, certainly the quantity of interest for the scattering analysis of
any mechanical theory of point particles and it seems
appropriate to {\it define} $\sigma^\psi$ in (\ref{SIGMA0})
as the cross section measure (see also the last section).

How can we find a handy expression for this probability?

With formula (\ref{EXPECT}) we have already a
formula for the expected number of  crossings
and the expected number of
signed crossings
of the surface $R\Sigma$  in the time interval $\Delta$.
For large $R$ the sphere $\PKR$ should be  crossed at most once, from
the inside
to the outside, such that the number of crossings equals the number of
signed crossings, both being either 0 or 1, such that furthermore their
expectation value  equals the probability, that the particle has crossed
the surface $R\Sigma$ at some time.

Hence, if there are asymptotically no backward crossings, i.e.\
 if $\lir \IE^\psi(N_-^{\PKR,[0,\infty)} ) = 0$,
we find for the asymptotic {\it probability}
that a trajectory crosses the surface $R\Sigma$ from the inside to the
outside
\ba
\label{SIGMA1}
\sigma(\Sigma ) =
\lir \IP^\psi(\bx_e\in R\Sigma) &=& \lir \int_0^\infty dt \int_{R\Sigma}
\bj^{\psi_t} \cdot \bn d\sigma.
\ea
This is a very nice result, because it connects our (natural)
definition of the cross section (\ref{SIGMA0}) with the quantity
considered in the \FAS\ (\ref{FAST}).

Up to now  the discussion leading to  formula
(\ref{SIGMA1}) has been completely general concerning
the time evolution.
Let us now process  (\ref{SIGMA1}) further, taking the
 simplest case, namely free evolution.  Our goal is to find a formula
where
the limit is taken.

The flux will contribute to the integral in (\ref{SIGMA1}) only for large
times,
because the packet has to travel a long time until it reaches the distant
sphere $\PKR$
such that we may use the long-time asymptotics of the free evolution.
We use the well-known formula (Reed, Simon 1975, p.59)
\be
\psi_t(\bx) = (e^{-iH_0t} \psi)(\bx) =
\int d^3y \frac{ e^{i\frac{|\bx-\by|^2}{2t}} }{(2\pi it)^{3/2}} \psi(\by)
\ee
and obtain with the splitting
\be
\psi_t(\bx) = \frac{e^{i\xxt}}{(it)^{3/2}} \hat{\psi}(\xt)
+ \frac{e^{i\xxt}}{(it)^{3/2}} \int \frac{d^3y}{(2\pi)^{3/2}}e^{-i\xyt}
(e^{i\yyt} - 1) \psi(\by),
\ee
neglecting the second term, as $t\to\infty$
\be
\label{ASYMP}
\psi_t(\bx) \sim (it)^{-3/2}e^{i\xxt} \hat{\psi}(\xt).
\ee
This asymptotics
for scattering theory has since long been realized as important (e.g.\
Brenig and  Haag 1959 and
 Dollard 1969, who proved that the asymptotics (\ref{ASYMP})
holds in the $L_2$ sense).
{}From
(\ref{ASYMP}) we find for  $t\to\infty$
\be
\label{JASYMPHEU}
\jpsi(\bx) = {\mbox{\rm Im}}  \/ \psi_t^*(\bx) \nabla \psi_t(\bx) \approx
\xt t^{-3} |\hat{\psi}(\xt)|^2,
\ee
and note, that the current and hence also the velocity
is strictly radial for large times, i.e.\
parallel to the outward normal $\bn$ of $\PKR$, reflecting our
expectation that the
expected value of backward
crossings of $\PKR$  vanishes as $R\to\infty$.

Back to  (\ref{SIGMA1}). Using the approximation
(\ref{JASYMPHEU})  and substituting $\bv := \xt$ we arrive
at
\ba
\label{NICE}
\int_0^\infty dt \int_{R\Sigma} \jpsi \cdot \bn d\sigma
&\approx&
\int_0^\infty dt \int_{R\Sigma} t^{-3}|\psihat(\xt)|^2  \xt \cdot
\bn(\bx) d\sigma \nonumber \\
&=& \int_0^\infty dv v^2 \int_C d\Omega |\psihat(\bv)|^2 =
\int_C d^3v |\psihat(\bv)|^2.
\ea
This heuristic argument for the free \FAS\ (\ref{FASFREET})
is so simple and intuitive that it is indeed
strange that it does not appear in any primer on scattering theory!

Now let us turn to the interacting case. Using Ikebe's eigenfunction
expansion
(\ref{PSIT}) and the relation of the generalized eigenfunctions with
the wave operators we
find that
\be
\psi_t(\bx) =  (2\pi)^{-3/2}\int d^3k e^{-i\kkt}  \phi(\bx,\bk)
\widehat{\Omega_-^\dagger\psi}(\bk).
\ee
The Lippmann-Schwinger equation (\ref{LS1}) for $\phi(\bx,\bk)$
allows us
to split off the free evolution of $\Omega_-^\dagger\psi$ and we
obtain
\ba
\psi_t(\bx) &=& e^{-iH_0t}\Omega_-^\dagger\psi \nonumber \\
&-&
(2\pi)^{-3/2}\int d^3k e^{-i\kkt} \widehat{\Omega_-^\dagger\psi}(\bk)
\Bigl(\frac{1}{2\pi} \int d^3y \frac{e^{-i{k|\bx-\by|}}}{|\bx-\by|}
V(\by) \phi(\by,\bk)\Bigr).
\ea
Now observe that the first term immediately gives with (\ref{NICE})
the desired result
\be
\label{INTFORM}
\lir \int_0^\infty dt \int_{C\cap\PKR} \bj^{\psi_t} \cdot \bn d\sigma=
\int_C d^3v |\widehat{\Omega_-^\dagger\psi}(\bv)|^2.
\ee
The other terms which appear in the flux
should not contribute for the following reason: they all contain the
phase factor  $e^{-i\kkt-ikx}$, where we used $e^{-ik|\bx-\by|}
\approx e^{-ikx}$
for $x$ far away from the range of the potential.
 This factor is rapidly oscillating
for large
$x$ and $t$ such that  the $k$-integrals
should decay fast enough in $x$
and $t$ to give no contribution to the flux across surfaces.

\section{Morals}

Let us recollect what we have achieved from our Bohmian perspective.

The analysis of the scattering regime of \bm\ suggests a natural
{\it definition}
of the cross section measure as the asymptotic probability distribution
of the
exit positions, formula (\ref{SIGMA0}).
This
is very similar to the definition of the cross section in classical
mechanics
(e.g.\ Reed, Simon 1979, p.15). The random distribution of impact
parameters
used in classical mechanics to define the cross section measure
corresponds to
the $|\psi|^2$ distribution of the initial positions of the particle.
The ``individual event'' (cf.\ the Born quote at the beginning),
that is the deflection of one particle in a certain direction,
is indeed determined by the initial position. However, the
initial positions are randomly distributed according to $|\psi|^2$
such that
there is no way to control the individual event. What is relevant
for
a scattering  experiment with a given wave function
is the statistical distribution of these individual
events and the  \FAS\ provides us with a formula for the asymptotic
distribution.

Let us now examine how this formula may be connected with the usual
operator formalism of quantum mechanics.
The first step has already been done, by writing formula (\ref{SIGMA0})
in terms of wave operators instead of using the
generalized eigenfunctions directly. (Existence of the wave operators
and asymptotic
completeness may
appear as a by-product, once an eigenfunction expansion has been
established
(Green and Lanford 1960).)
Note that formula (\ref{SIGMA0}) may be rewritten as
\be
\sigma(\Sigma) = \int_C d^3v |\widehat{\Omega_-^\dagger\psi}(\bv)|^2
= (\psi, \Omega_-{\cal F}^{-1}P_C{\cal F} \Omega_-^\dagger \psi),
\ee
where $P_C$ denotes the projection operator on the cone $C$ and
${\cal F}$ is
the unitary operator of the fourier transformation. The map
$\Sigma \mapsto
 \Omega_-{\cal F}^{-1}P_C{\cal F} \Omega_-^\dagger$ is an
explicit example
of what is called a  projection operator valued measure which
corresponds to a unique
\sa\ operator by the spectral theorem. This particular example
of escape statistics
exemplifies the general situation, namely that operators as
``observables'' appear
merely as computational tools in the phenomenology of certain
types of experiments---those
for which the statistics of the result are governed by a projection
operator valued measure.

Now we come to the end of the closing remarks of the Born quote
at the
beginning. Is it true that any deterministic completion of
quantum mechanics must lead
to the
same formulae for the cross section as Born's formula?

Certainly not!
Further assumptions would be required, e.g.\
the initial packet must be close to a plane wave, initially far away
from the scattering
center, in order to have a chance to arrive from  formula (\ref{FAST})
(or from  Dollard's formula (\ref{SICT}))
at Born's formula (\ref{BORNFORM}).

This is however not the strongest
point we can make here, because
Born's formula is not really taken seriously as the fundamental formula
for the cross section nowadays; but Dollard's formula certainly is.

But what is the quantum mechanical prediction if we go further and
do not want to make
such strong idealizations, e.g.\ if we happen to
place detectors around the scattering center
which are {\it not}\/ that
far away? What are the
predictions for the times and positions at which the detectors click?
Dollard's formula doesn't apply here, because it is an asymptotic formula.

Maybe---after some reflection---one comes up with what seems a very natural
candidate for the joint density of exit position and exit time, namely
$\jpsi(\bx) \cdot {\bf n}({\bx})
 d\sigma dt$.
But further
scrutinizing this answer within quantum mechanics should leave one uneasy.
After all,  this is a formula---a prediction---which is not at all
a quantum mechanical prediction of the common type:
 $\jpsi(\bx) \cdot {\bf n}({\bx})$ can well be negative and thus
the formula makes  sense only for (very) particular wave functions,
for which
$\jpsi(\bx) \cdot {\bf n}({\bx})$ is positive.
But what is then the right formula for
all the other wave functions? Why is there such a simple formula for some
wave functions?

In \bm\ there is no need to feel uneasy. $\IP^\psi(({\bx}_e,t_e) \in
(d\sigma,dt))$ is the
probability
that the particle exits at $\bx_e$ at time $t_e$---and that
can be calculated in the usual way for {\it any} \wf .
For some \wf s it turns out  (Daumer, D\"urr, Goldstein and Zangh\a'{\i}
1994) to be
indeed given by
\be
\label{FUNDAMENT}
\IP^\psi(({\bx}_e,t_e) \in (d\sigma,dt)) = \jpsi(\bx) \cdot {\bf n}({\bx})
 d\sigma dt,
\ee
for others it isn't.
That's alright, isn't it?

\newpage

\begin{figure}[b]
\label{STREUNG}
\begin{center}
\leavevmode%
\epsfxsize=13cm
\epsffile{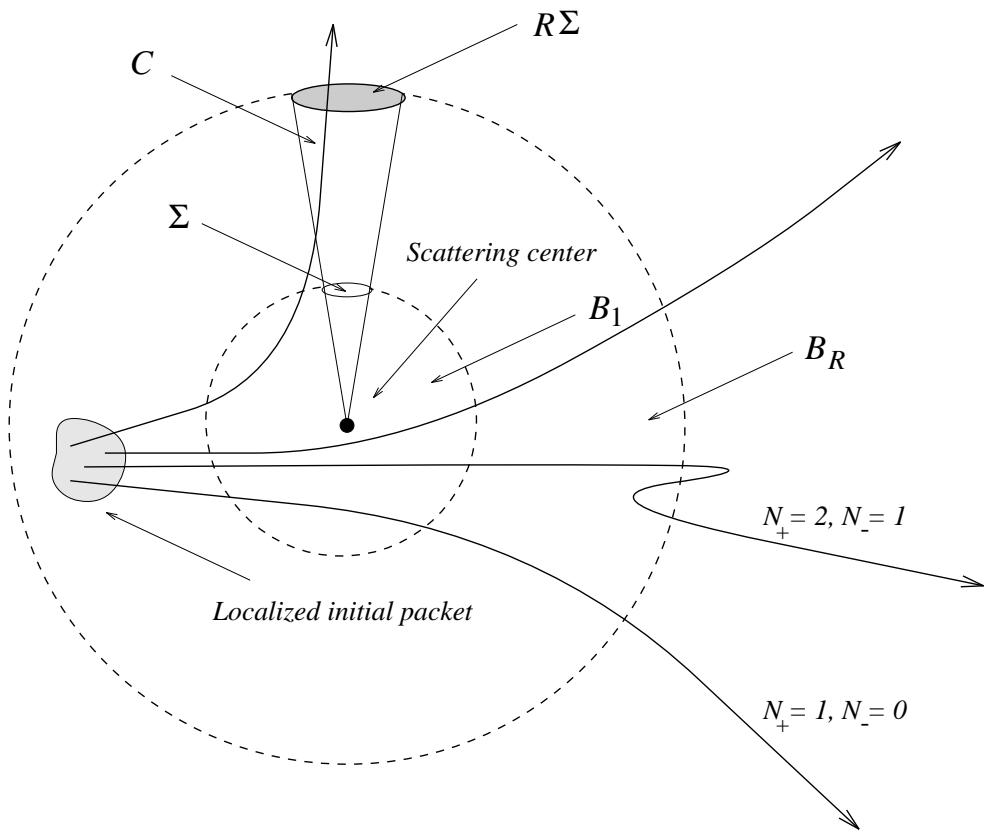}
\end{center}
\caption{Possible trajectories of a particle
starting in the wave packet localized in the ball $B_R$.}
 \end{figure}

\newpage

\section*{References}
\setlength{\parindent}{0pt}
\parskip 3ex plus 0.1ex minus 0.1ex

Bell, J.S. (1987), {\it Speakable and Unspeakable in Quantum
Mechanics}, Cambridge University Press.

Beller, M. (1990), ``Born's probabilistic interpretation: a case
study of `concepts in flux',''
{\it Stud.~Hist.~Phil.~Sci.} {\bf 21}, 563-588.

Bohm, D. (1952), ``A suggested interpretation of the quantum theory in terms
of ``hidden'' variables I, II,'' {\em Phys.~Rev.} {\bf 85}, 166--179,
180--193.

Bohm, D., Hiley, B.J. (1993), {\em The Undivided Universe:
An Ontological Interpretation  of Quantum Theory}, Routledge \& Kegan Paul,
London.

Berndl, K. (1995),  {\it Zur Existenz der Dynamik in Bohmschen Systemen,}
PhD thesis, Mainz Verlag, Aachen.

Born, M. (1926), ``Quantenmechanik der Sto\ss vorg\"ange,'' {\em Zeitschrift
f\"ur Physik}
 {\bf 38}, 803--827, (translated in Ludwig (1968), p.\ 94-105).

Brenig, W., Haag, R. (1959), ``General quantum theory of collision
processes,''
{\em Fortschr.~d.~Phys.}
{\bf 7}, 183--242, (translated in Ross (1963), 14--108).

Combes, J.-M., Newton, R.G.,  Shtokhamer, R. (1975),
``Scattering into cones and flux across surfaces,''
{\em Phys.~Rev.~D} {\bf 11},  366--372.

Daumer, M. (1995),  {\it Streutheorie aus der Sicht Bohmscher Mechanik,}
PhD thesis, Mainz Verlag, Aachen.

Daumer, M., D\"urr, D., Goldstein, S., Zangh\a'{\i}, N. (1994),
``Scattering and the
role of operators in Bohmian mechanics,'' 331--338, {\it On Three Levels},
M.~Fannes,
C.~Maes, and A.~Verbeure (eds.),
NATO ASI series 324, Plenum, New York.

Daumer, M., D\"urr, D., Goldstein, S., Zangh\a'{\i}, N. (1995),
``On the flux across surfaces theorem'',
in preparation.

Derezi\'{n}ski, J., Gerard, C. (1993), {\it Asymptotic Completeness
of $N$-particle Systems}, Preprint, Erwin Schr\"odinger International
Institute for
Mathematical Physics, ESI 60.

Dollard, J.D. (1964), ``Asymptotic convergence and the Coulomb
interaction,''
{\em J.~Math.~Phys.} {\bf 5}, 729--738.

Dollard, J.D. (1969),  ``Scattering into cones I, Potential scattering,''
{\em Comm.~Math.~Phys.} {\bf 12},  193--203.

D\"urr, D., Goldstein, S., Zangh\a'{\i}, N. (1992),
``Quantum equilibrium and
the origin of absolute uncertainty,'' {\em J.~Stat.~Phys.}, {\bf 67}, 843--907.

Enss, V., Simon, B.,  (1980), ``Finite total cross section in
nonrelativistic quantum mechanics,''
{\it Comm.\ Math.\ Phys.} {\bf 76}, 177-209.

Enss, V. (1978), ``Asymptotic completeness for quantum mechanical
potential scattering,'' {\em Comm.~Math.~Phys.} {\bf 61}, 285.

Faddeev, L. (1963), {\it Mathematical aspects of the three-body
problem in quantum scattering theory,} (Steklov Math.\ Inst.\ {\bf 69}),
Israel
Program for Scientific Translation.

Gell-Mann, M., Goldberger, (1953), ``The formal theory of scattering,''
{\it Phys.\ Rev.} {\bf 91}, 298--408.

Green, T.A., Lanford III, O.E. (1960), {\em J.~Math.~Phys.} {\bf 1}, 139.

Heisenberg, W. (1925), ``\"Uber die quantentheoretische Umdeutung
kinematischer und
mechanischer Beziehungen,'' {\em Z.~f.~Physik} {\bf 33}, 879.

Heisenberg, W. (1946), ``Der mathematische Rahmen der Quantentheorie
der Wellenfelder,''
{\em Z.~f.~Naturforschung} {\bf 1}, 608--622.

Holland, P. (1993), {\it The quantum theory of motion},
Cambridge University
press.

Ikebe, T. (1960), ``Eigenfunction expansions associated with the
Schr\"odinger operator
and their application to scattering theory,'' {\em Arch.~Rational
Mech.~Anal.} {\bf 5}, 1--34.

Leavens, C.R. (1990), ``Transmission, reflection and dwell times
within Bohm's causal interpretation of quantum mechanics,''
{\em Solid State Comm.} {\bf 74}, 923.

Lippmann, B., Schwinger, J. (1950), ``Variational principles
for scattering processes I,''
{\em Phys.~Rev.} {\bf 79}, 469--480.

Ludwig, G. (1968), {\it Wave mechanics}, Pergamon Press, Oxford.

M{\o}ller, C. (1946), {\em Kgl. Danske Videnskab. Selskab.,
Mat.~-Fys.~Medd.} {\bf 23}, 1,
(reprinted in Ross (1963), p.109-154).

Reed, M, Simon, B. (1975), {\it  Methods of Modern Mathematical
Physics II}, Academic Press Inc., London.

Reed, M, Simon, B. (1979), {\it  Methods of Modern Mathematical
Physics  III}, Academic Press Inc., London.

Ross, M. (1963), {\it Quantum scattering theory}, Indiana
University Press, Bloomington.

Ruelle, D. (1969), ``A remark on bound states in potential scattering
theory,'' {\em Nuovo Cimento A} {\bf 61}, 655.

Schr\"odinger, E. (1926), ``Quantisierung als Eigenwertproblem,''
{\em Ann.~Phys.} {\bf 79}, 361--376, (translated in Ludwig (1968),
p.94--105).

Simon, B. (1971),  {\it Quantum Mechanics for Hamiltonians
defined as Quadratic Forms},
Princeton University Press, Princeton, New Jersey.

\end{document}